\documentstyle[graphicx]{elsart}

\newcommand{\beq}{\begin{equation}}
\newcommand{\eeq}{\end{equation}}
\newcommand{\bqa}{\begin{eqnarray}}
\newcommand{\eqa}{\end{eqnarray}}

\newcommand{\bld}[1]{\mbox{\boldmath$\mathrm{#1}$}}
\newcommand{ \bb }{$2\nu\beta\beta$}
\newcommand{ \bbm }{$2\nu\beta^-\beta^-$}

\begin{document}
\begin{frontmatter}
\title{Sensitive behavior of \bb-decay amplitude within QRPA and broken SU(4) symmetry in nuclei}

\author[TUEB]{ Vadim A.~Rodin \thanksref{e-mail}},
\author[MEPI,KVI]{Michael H.~Urin} and
\author[TUEB]{Amand Faessler}

\address[TUEB]{Institut f\"ur Theoretische Physik der Universit\"at T\"ubingen,
Auf der Morgenstelle 14, D-72076 T\"ubingen, Germany}
\address[MEPI]{Department of Theoretical Nuclear Physics,
Moscow State Engineering and Physics Institute,115409 Moscow, Russia}
\address[KVI]{Kernfysisch Versneller Institute, NL-9747AA Groningen, The Netherlands}
\thanks[e-mail]{E-mail: vadim.rodin@uni-tuebingen.de}

\begin{abstract}
Making use of an identity transformation independent of a nuclear model, 
we represent the {\bb}-amplitude as a sum of two terms.
One term accounts for most of 
the sensitivity of the original {\bb}-amplitude to $g'_{pp}$ for 
realistic $g'_{pp}\simeq 1$ (with $g'_{pp}$ being the ratio of the triplet 
and singlet p-p interaction strengths) and
is determined by a specific energy-weighted sum rule. The sum rule depends only on the
particle-particle residual interaction (being linear function of $g'_{pp}$ in the QRPA) and 
passes through zero at the point $g'_{pp}=1$ where the Wigner SU(4) symmetry is restored 
in the p-p sector of the Hamiltonian. The second term in the decomposition of
the {\bb}-amplitude is demonstrated within the QRPA to be a much smoother function 
for the realistic values of $g'_{pp}$ than the original {\bb}-amplitude. This term is mainly determined by the intensity 
of the spin-orbit interaction of the nuclear mean field.
Thus, the analysis of the present work reveals the reasons for 
the sensitivity of the {\bb}-amplitude to different components of the nuclear Hamiltonian and thereby can help in constraining nuclear model uncertainties in calculations of the amplitude.

\begin{flushleft}
\end{flushleft}
\begin{flushleft}
\end{flushleft}

\end{abstract}
\end{frontmatter}
\section{Introduction}
Numerous calculations of the \bb-amplitude within the QRPA have revealed its sensitive dependence
on the strength $g_{pp}$ of the charge-exchange triplet particle-particle (p-p) interaction 
(see, e.g., reviews~\cite{Fae98}).
The amplitude passes through zero at a realistic value of the interaction strength.
However, the mechanism of such a dependence found for the first time in the pioneering work~\cite{Vog86}
almost 20 years ago still remains unexplained. The paper~\cite{Vog86} already contained a clue that the vanishing
of the amplitude may be related to the restoration of the Wigner spin-isospin SU(4) symmetry in nuclei.
In an exactly solvable schematic model the authors of~\cite{Vog86} also showed that this kind of dependence
is not an artifact of the QRPA but shows up in the exact solution.

There have been several semiquantitative attempts to use the concept of the dynamically broken SU(4) symmetry
as a basis for describing the \bb-amplitude~\cite{Krmp92}.
Since the amplitude is zero when the SU(4) symmetry is exact, it seems
natural to express the amplitude explicitly in terms of those parts of the nuclear Hamiltonian $\hat H$ which are responsible
for the violation of the symmetry. To achieve this goal, a model-independent transformation of the amplitude was
suggested in~\cite{Rum98}. Unfortunately, the quantitative realization
of this idea in~\cite{Rum98} was based on a simplified
model of independent quasiparticles and, therefore, could not address the question of
the sensitivity of the \bb-amplitude calculated within the QRPA. That was the reason why
the work~\cite{Rum98} faced strong criticism claiming that ``the formalism fails in describing
the double beta decay observables"~\cite{Civ00}.

To refute the criticism of~\cite{Civ00}, in present work we substantially develop the approach~\cite{Rum98}
by incorporating the QRPA into it.
As in~\cite{Rum98}, the starting point for the analysis is model-independent, identity transformation
of the \bb-amplitude: $M_{2\nu}=M_{2\nu}'+S/\omega_g^2$. 
That allows us to separate out a part proportional to an specific energy-weighted sum rule 
$S$ depending only on the particle-particle residual interaction. This part is shown within the QRPA to account for most of the sensitivity of the original {\bb}-amplitude to $g'_{pp}$ for realistic
$g'_{pp}\simeq 1$ (with $g'_{pp}$ being the ratio of the triplet and singlet p-p interaction strengths).
The sum rule $S$ can be expressed in the form that avoids the intermediate state summation and involves 
only the wave functions of the initial and final ground states. 
An analytical representation for $S$ that can be given within the QRPA using a simple separable p-p interaction
reveals its simple linear dependence on $g'_{pp}$ with passing through zero at $g'_{pp}=1$ where the SU(4) symmetry is restored in the particle-particle sector of the Hamiltonian. The second part $M_{2\nu}'$ in the decomposition of the \bb-amplitude can be explicitly expressed in terms of those components 
of the nuclear Hamiltonian which violate the SU(4) symmetry. 
As shown by our QRPA calculations, this part behaves much smoother for the realistic values of 
$g'_{pp}$ 
and is mainly determined by the spin-orbit part of the nuclear mean field
as was conjectured by Rumyantsev and Urin~\cite{Rum98}.

Thus, we think that our present results contribute to understanding of general physical properties of the \bb-decay
amplitude and bring new opportunity to constrain nuclear model uncertainties in calculations of the amplitude.
In particular, it is shown that the sensitivity of the amplitude to the p-p interaction strength has
its natural physical origin in the
violation of the SU(4) symmetry in the particle-particle sector of a realistic nuclear Hamiltonian
and has nothing to do with the shortcomings of the QRPA itself.
Also an important influence of the spin-orbit mean field on the calculation results
that has received rather little attention in the preceding publications is emphasized.
Besides, the sum rule $S$ can be easily calculated in different nuclear models, for instance,
in shell model, thereby providing a test of the quality of the QRPA calcuations.

Note, that we do not claim (as well as the authors of~\cite{Rum98} did not) any
{\em model-independent calculations} of the \bb-amplitude (that was misunderstood by
authors of~\cite{Civ00}) rather we are dealing with {\em model-independent transformation}
of the \bb-amplitude. Nevertheless, as shown in the paper this identity
transformation allows us to shed some light on the general properties of the amplitude
calculated within the QRPA.

The paper is organized as follows. In section~2 an identity transformation of the \bb-amplitude
which is valid independently of a nuclear structure model is performed.
The transformation turns out to be a power tool for the analysis of the
\bb-decay amplitude because it allows one to incorporate explicitly
the symmetry properties of the nuclear Hamiltonian. The model Hamiltonian consisting
of a mean field, the Landau--Migdal particle-hole interaction and a separable particle-particle
interaction is specified in section~3. The identity transformation separates out a part proportional to
a specific energy-weighted sum rule for which an analytical representation is given within the QRPA
in section~4. Calculation results for $^{76}$Ge, $^{100}$Mo and $^{130}$Te are presented
in section~5 and we conclude in section~6.

\section{Model-independent transformation of the \bb-amplitude}

We adopt here the same line of reasoning as presented in~\cite{Rum98}.
To make the following derivation of the new representation for the \bb-decay
amplitude more transparent, we will use the same approach for
the Fermi and the Gamow-Teller transitions.
Let $\hat G^{(\pm)}=\sum_a g_a^{(\pm)}$ be the operators of allowed Fermi
($g^{(\pm)}=\tau^{(\pm)}$) and GT ($g^{(\pm)}=\bld\sigma\tau^{(\pm)}$)
$\beta$ - transitions.

We start our derivation with writing down the usual expression for
the {\bbm}-decay amplitude (see, e.g., reviews~\cite{Fae98}):
\beq
\label{Mbb}
M_{2\nu} = \sum_s \frac{g_s f_s}{\omega_s}\ .
\eeq
Here,  $g_s=\langle f \vert \hat G^{(-)}\vert s\rangle$, $f_s=\langle s\vert
\hat G^{(-)} \vert i \rangle$ are the one-step matrix elements of the operator $\hat G^{(-)}$
between the intermediate states $\vert s\rangle$ of an isobaric nucleus $(N-1,Z+1)$
and the ground states
$\vert i \rangle$ and $\vert f\rangle$ of a parent ($Z,N$) and the final
($Z+2,N-2$) nucleus, respectively, with $E_s$, $E_i$ and $E_f$ being the corresponding energies
and $\omega_s = E_s-(E_i + E_f)/2$ being the excitation energy of an intermediate state
measured from the mean value of the ground-state energies.

Then we perform the following identical splitting of the amplitude (\ref{Mbb}) into two parts:
\bqa
& M_{2\nu}=M_{2\nu}'+\frac{S}{\omega_g^2} \label{Mbb1}\\
& M_{2\nu}' = \sum_s \frac{(\omega_g^2-\omega_s^2)g_s f_s}{\omega_g^2\omega_s} \label{Mbb'}\\
& S=\sum_s \omega_s g_s f_s, \label{Mbb2}
\eqa
where $\omega_g$ is an arbitrary energy
that was identified with the energy $\omega_G$ of the corresponding giant resonance in~\cite{Rum98}.
It should be so in the limit of exact SU(4) symmetry, but for the realistic situation one can only say
that $\omega_g$ should be rather close to $\omega_G$.
Note, that the second term in (\ref{Mbb1}) is proportional to $S$ having the form (\ref{Mbb2})
of a specific energy-weighted sum rule. Bearing in mind that
\bqa
\label{delta}
\nonumber \omega_s f_s
= \langle s\vert \left[\hat H,\hat G^{(-)}\right] \vert i\rangle
-(E_f - E_i) f_s /2
\ ,\\
-\omega_s g_s
={\langle f\vert \left[\hat H,\hat G^{(-)}\right]\vert s\rangle}-(E_f - E_i) g_s /2 \ ,
\eqa
where $\hat H$ is the nuclear Hamiltonian, one can get the following expression for $S$:
\beq
\label{S}
S = \frac{1}{2}\langle f\vert \left[\hat G^{(-)},\left[\hat H,\hat G^{(-)}\right]\right]\vert i\rangle\ .
\eeq
Here we do not support the claim of~\cite{Rum98} that $S$ should vanish (see Eq.~(5) of the citation). That would be true only if
the p-p interactions in the Hamiltonian could be neglected. But it has been known for a long
time that the p-p interactions play an important role in the nuclei far from closed shells,
especially for describing {\bb}-decay amplitudes~\cite{Vog86}. Actually, the authors of~\cite{Civ00} provided direct
evidence by their QRPA calculations that $S$ does not vanish (except for one point in $g_{pp}$ dependence, see Fig.~1 of
the reference).

We will show in Section 4 that one can get within the QRPA an analytical expression for $S$ depending linearly on
$g_{pp}$. The point where $S=0$ then corresponds to the restoration of the SU(4)-symmetry in the particle-particle
sector of the Hamiltonian (i.e. $\hat G^{(-)}$ commutes with the p-p part of the Hamiltonian for this $g_{pp}$).

To conclude this section, we show, following~\cite{Rum98}, that $M_{2\nu}'$ can be written as
\beq
M_{2\nu}' = \frac{1}{\omega_g^2}\sum_s \frac{\tilde g_s \tilde f_s}{\omega_s}\ ,
\label{M_G'}
\eeq
where $\tilde f_s={\langle s\vert \hat V_G^{(-)}\vert i\rangle}=-({\omega_g-\omega_s})\langle s\vert \hat G^{(-)}\vert i\rangle$ and
$\tilde g_s={\langle f\vert \hat V_G^{(-)}\vert s\rangle}=-({\omega_g+\omega_s})\langle f\vert \hat G^{(-)}\vert s\rangle$ are
the matrix elements of the operator $\hat V_G^{(-)}$ defined as:
\beq
\label{VG}
\hat V_G^{(-)}= \left[\hat H,\hat G^{(-)}\right] - \Delta_g \hat G^{(-)},\ \
\Delta_g = \omega_g +\frac{E_f- E_i}{2}\ .
\eeq
It is noteworthy that $M_{2\nu}'$ acquires contributions from all terms of the Hamiltonian violating the SU(4) symmetry
(in contrast to the sum rule $S$ (\ref{S}) which is determined only by the residual p-p interaction). On the mean field
level the most important source of the violation is the spin-orbit mean field as was conjectured by Rumyantsev
and Urin~\cite{Rum98}. We will see in Section 5 how strongly is $M_{2\nu}'$ affected by varying the intensity of
the spin-orbit mean field in the calculations.

Thus, the final expressions (\ref{S}) and (\ref{M_G'}),(\ref{VG}) contain explicit information about
symmetry properties of the nuclear Hamiltonian in terms of the corresponding commutators.
Also we would like to stress again that the results of this section ---
the transformation of the \bb-amplitude Eqs.(\ref{Mbb1})-(\ref{Mbb2}) and the representations
for $S$ and $M_{2\nu}'$  Eqs.(\ref{S})-(\ref{M_G'}) --- are quite general and do not rely
on any nuclear model.

\section{Model Hamiltonian}

For a quantitative realization of the relationships of the preceding section one has to
specify the nuclear Hamiltonian.
We adopt here the realistic model Hamiltonian of~\cite{Rod03}
consisting of a mean field $U(x)$, the Landau--Migdal zero-range forces for describing
the particle-hole interaction $\hat H_{p\mbox{\it-}h}$ and a separable particle-particle interaction
in both the neutral and the charge-exchange channels $\hat H_{p\mbox{\it-}p}$.
Only isoscalar (central $U_0(r)$ plus spin-orbit $U_{so}(r)\bld{\sigma l}$) part of the mean field $U(x)$ is treated
phenomenologically while the isovector (symmetry potential) and the Coulomb parts are calculated consistently in
the Hartree approximation (see~\cite{Rod03}).
The explicit expression for the Landau--Migdal forces in the charge-exchange channel is~\cite{Mig83}:
\beq
\hat H_{p\mbox{\it-}h}=2\sum\limits_{a>b} (F_0+F_1\bld\sigma_a\bld\sigma_b)\tau^{(-)}_a\tau^{(+)}_b
\delta (\bld r_a - \bld r_b) + h.c.\label{Hp-h}
\eeq
where the intensities of the non-spin-flip and spin-flip parts of this
interaction, $F_0$ and $F_1$ respectively, are the phenomenological parameters.
We choose the p-p
interaction in the following separable form (as used in the BCS model)
in both the neutral (pairing) and charge-exchange particle-particle channels:
\beq
\hat H_{p\mbox{\it-}p}=-\frac{1}{2}\sum\limits_{J\mu}G_J\!\!\!\!
\sum\limits_{\scriptstyle\beta=p,n\atop\scriptstyle\beta'=p,n}\!\!
(\hat P^{J\mu}_{\beta\beta'})^+ \hat P^{J\mu}_{\beta\beta'}\ ,\label{2.1}
\eeq
where $G_{J}$ are the intensities of the particle-particle interaction.
Here we consider only the nucleon pairs with total angular momentum and parity $J^\pi=0^+,1^+\ (L=0, S=J)$.
The annihilation operator for a nucleon pair $P^{J\mu}$ is given as follows:
\bqa
&&\hskip-1cm \hat P^{J\mu}_{\beta\beta'}=\sum\limits_{\lambda\lambda'}
(\chi_\lambda^\beta \chi_{\lambda'}^{\beta'})
P^{J\mu}_{\beta\beta',\lambda\lambda'}\ ;\
P^{J\mu}_{\beta\beta',\lambda\lambda'}=
t^{(J)}_{(\lambda)(\lambda')}\sum\limits_{mm'}
(J\mu|j_\lambda^\beta m j_{\lambda'}^{\beta'} m') a^{\beta'}_{\lambda' m'}
a^{\beta}_{\lambda m}\ ,\label{2.2}
\eqa
where $a_{\lambda m}(a_{\lambda m}^+)$ is the annihilation (creation)
operator of a nucleon in the state with the quantum numbers $\lambda m$
($m$ is the projection of the particle angular momentum) and the single-particle
radial wave functions $r^{-1}\chi_{\lambda}(r)$, $(\chi_\pi\chi_\nu)=\int \chi_\pi(r)\chi_\nu(r)\,dr$,
$t^{(0)}_{(\pi)(\nu)}=\sqrt{2j_\pi+1}\delta_{(\pi)(\nu)}$,
$t^{(1)}_{(\pi)(\nu)}=\frac{1}{\sqrt 3}\langle(\pi)\|\sigma\|(\nu)\rangle$.

Note that, apart from the p-p interaction, the model Hamiltonian has proven its reliability
by successful continuum-RPA calculations of many structure and decay properties of
charge-exchange giant resonances (see, e.g.,~\cite{Rod}).

We use the Bogolyubov transformation to describe the nucleon pairing in the
neutral channels in terms of quasiparticle creation (annihilation) operators
$\alpha_{\lambda m}^+(\alpha_{\lambda m})$ (see, e.g.,~\cite{Sol}). As a
result, the following model Hamiltonian is obtained to describe Fermi and Gamow-Teller excitations
within the pn-QRPA:
\beq
\hat H=\hat H_{0}+\hat H_{p\mbox{\it-}h}+\hat H_{p\mbox{\it-}p}+\mu_n \hat N + \mu_p \hat Z ;\
\hat H_{0}=\sum\limits_{\beta=p,n}\sum\limits_{\lambda m}
E^\beta_\lambda (\alpha^\beta_{\lambda m})^+\alpha^\beta_{\lambda m}
\label{2.4}
\eeq
Here, $E^\beta_\lambda=\sqrt{(\xi^\beta_\lambda)^2+\Delta_\beta^2}$
is the quasiparticle energy, $\xi^\beta_\lambda=
\varepsilon^\beta_\lambda-\mu_\beta$, $\mu_\beta$ and $\Delta_\beta$
are the chemical potential and the energy gap, respectively,
which are determined from the BCS-model equations:
\beq
N_\beta=\sum_\lambda
(2j^\beta_\lambda+1) (v^\beta_\lambda)^2;\ \ \
\Delta_\beta=G_0 \sum_\lambda
(2j^\beta_\lambda+1) u^\beta_\lambda v^\beta_\lambda
\label{2.5}
\eeq
with $v^2_\lambda=\frac{1}{2}(1-\frac{\xi_\lambda}{E_\lambda})$
and $u^2_\lambda=1-v^2_\lambda$.

The explicit expressions for the p-h and p-p interactions in terms of
the quasiparticle (pn)-pair creation and annihilation operators are:
\bqa
&&\hat H_{p\mbox{\it-}h}=\sum\limits_{J\mu}\frac{2F_J}{4\pi}
\sum_{\pi\nu\pi'\nu'}(\chi_\pi \chi_\nu \chi_{\pi'} \chi_{\nu'})
(Q^{J\mu}_{\pi\nu})^+Q^{J\mu}_{\pi'\nu'},\label{2.7}\\
&&\hat H_{p\mbox{\it-}p}=-\sum\limits_{J\mu}G_J\sum_{\pi\nu\pi'\nu'}
(\chi_\pi \chi_\nu)(\chi_{\pi'} \chi_{\nu'})
(P^{J\mu}_{\pi\nu})^+P^{J\mu}_{\pi'\nu'},\label{2.8}\\
&&\hskip-1cm Q^{J\mu}_{\pi\nu}=t^{(J)}_{(\pi)(\nu)}(u_\pi v_\nu A^{J\mu}_{\pi\nu}+
v_\pi u_\nu (\tilde A^{J\mu}_{\pi\nu})^+),
P^{J\mu}_{\pi\nu}=t^{(J)}_{(\pi)(\nu)}(u_\pi u_\nu A^{J\mu}_{\pi\nu}
-v_\pi v_\nu (\tilde A^{J\mu}_{\pi\nu})^+), \nonumber\\
&& \nonumber
\eqa
where $A^{J\mu}_{\pi\nu}=\sum\limits_{mm'}
(J\mu|j_\pi mj_{\nu} m') \alpha_{\nu m'} \alpha_{\pi m}$,
$\tilde A^{J\mu}_{\pi\nu}=(-1)^{J+\mu} A^{J\;-\mu}_{\pi\nu}$,
$(\chi_\pi \chi_\nu \chi_{\pi'} \chi_{\nu'})=
\int \chi_\pi \chi_\nu \chi_{\pi'} \chi_{\nu'} r^{-2}dr$. Note, that the radial integrals
in the representation (\ref{2.8}) for the separable p-p interaction should be kept
rather than put equal to 1 as in many applications of the separable forces. This is due to
a slight non-orthogonality of proton and neutron wave functions having the same quantum numbers.

Our model Hamiltonian preserves the isospin symmetry provided that a selfconsistency condition relating
the isovector part of the mean field (symmetry potential) to the Landau--Migdal parameter $F_0$ is fulfilled
(see~\cite{Rod03}, the p-p interaction (\ref{2.1}) is isospin invariant by construction). That was proved
by the direct continuum-QRPA calculations of the properties of the isobaric analog states in~\cite{Rod03}.
Note, that the usual discretized version of the pn-QRPA which we are going to exploit here violates a bit the isospin symmetry
due to a restricted basis of the particle-hole configurations. Apart from this source of the violation, we shall
introduce another, usual one, namely, we allow the pairing constants for proton and neutron subsystems ($G_{0p}$ and $G_{0n}$, respectively)
to have different values in order to reproduce the experimental pairing gaps. These approximations certainly affect the quality
of the calculated Fermi matrix elements that is worth to discuss elsewhere. However, they seem to be of rather
minor importance for the description of the Gamow-Teller matrix elements that we are pursuing here.

\section{Analytical representation for $S$}

In this section we show that a simple and elegant analytical representation for
the EWSR $S$ (\ref{Mbb2}) introduced in Section 2 can be obtained using
the Hamiltonian of the preceding section.
To do that, one has to calculate the expectation value of the double commutator (\ref{S}).
The equation of motion for the GT operator
can be derived within the QRPA~\cite{Rum98,Rod03}:
\bqa
&[\hat H, \hat G^{(-)}_\mu]=\left(\hat U^{(-)}_\mu\right)_{ph}+\left(\hat U^{(-)}_\mu\right)_{pp},
\label{1.6}\\
&\left(\hat U^{(-)}_{\mu}\right)_{pp}=
\frac{G_{0n}-G_1}{G_{0n}}\Delta_n (\hat P^{1\mu}_{pn})^+ -
\frac{G_{0p}-G_1}{G_{0p}}\Delta_p \tilde{\hat P}^{1\mu}_{pn}\ .\label{2.11}
\label{1.6'}
\eqa
The rhs of (\ref{1.6}) is the sum of particle-hole $\left(\hat U^{(-)}_{\mu}\right)_{ph}$
and particle-particle $\left(\hat U^{(-)}_{\mu}\right)_{pp}$ operators
originating from the commutator of $\hat G^{(-)}$ with the p-h part (including the mean field) and
the p-p part of the Hamiltonian, respectively.
The explicit representation for $\left(\hat U^{(-)}_{\mu}\right)_{ph}$ can be found in~\cite{Rum98,Rod03}
but it is irrelevant for the current derivation of $S$ since $\langle f\vert \left[\hat G^{(-)},
\left(\hat U^{(-)}_{\mu}\right)_{ph}\right]\vert i\rangle=0$ due to $\tau^{(-)2}=0$.
Hence, only p-p part of the Hamiltonian contribute to $S$:
\beq
S=\langle f\vert \left[\hat G^{(-)},\left(\hat U^{(-)}\right)_{pp}\right]\vert i\rangle=
3(1-g'_{pp})\frac{\Delta_n\Delta_p}{G_{0n}G_{0p}}\cdot\frac{G_{0n}+G_{0p}}{2},
\label{Sfin}
\eeq
where we have introduced the parameter $g'_{pp}=2G_1/(G_{0n}+G_{0p})$ describing the relative
strength of p-p interaction in the triplet spin channel with respect to the singlet 
(pairing) one (not to be confused
with the parameter $g_{pp}$ used in the calculations with realistic nucleon-nucleon interactions to renormalize
the p-p part of them). In getting (\ref{Sfin}) we have used an approximation that the BCS vacua are the same
for the initial and final ground states. For the Fermi transitions factor 3 in (\ref{Sfin}) should be omitted and
$g'_{pp}=2G_0/(G_{0n}+G_{0p})$.

In the case of the restored isospin symmetry in the p-p channel, i.e. $G_{0n}=G_{0p}=G_0$,
the expression (\ref{Sfin}) can be simplified further:
\beq
S=3(1-g'_{pp})\frac{\Delta_n\Delta_p}{G_{0}}.
\label{Sfin1}
\eeq
The linear dependence of the sum rule $S$ on $g'_{pp}$ with crossing zero at the point of the restoration of SU(4)-symmetry ($g'_{pp}=1$)
in the p-p channel is of great importance. It shows analytically that at least
part of the \bb-decay amplitude reveals a quite general
sensitivity to $g'_{pp}$ which is not really restricted to the used approximations like the QRPA and BCS.
In fact, any realistic nuclear Hamiltonian should contain singlet and triplet p-p interactions
and any model of the ground states of open-shell nuclei should account for the pairing effects,
in particular, the increased probability to pick up a pair of nucleons in the $0^+$ state which is described
by $\Delta$ in the BCS model. Therefore, we believe that $S$ will be linearly decreasing with respect to $g'_{pp}$
in any realistic model (including modern shell models) while the slope of the dependence and its zero
may be slightly different from those given by (\ref{Sfin}),(\ref{Sfin1}).~\footnote{This our conjecture
has been confirmed by direct shell-model calculations of an anonymous Referee to whom we are indebted.}

\section{Calculation results}
To illustrate the method suggested, we present in this section some calculation results
obtained within the pn-QRPA for $^{76}$Ge, $^{100}$Mo and $^{130}$Te.
Parameterization of the isoscalar part of the mean field $U_0(x)$ along with
the values of the parameters used in the calculations have been described
in details in~\cite{Gor01b}. Here we increase the depth of the central part $U_0(r)$ by about 10 MeV
to get a large enough basis of the bound single-particle states (this change of a isoscalar parameter should not
affect much the isovector quantities like the Gamow-Teller matrix elements). The dimensionless intensities $f_J$ of
the Landau-Migdal forces of Eq.~(\ref{Hp-h}) are chosen as usual: $F_J=f_J\cdot300$ MeV\,fm$^3$.
The value $f_0=f'=1.0$ of the parameter $f'$ determining the symmetry potential
according to the selfconsistency condition is also taken from~\cite{Gor01b}, where the
experimental nucleon separation energies have been satisfactorily described
for closed-shell subsystems for a number of nuclei.
The dimensionless intensity $f_1=g'$ of the spin-isovector part of the Landau-Migdal forces
is chosen as $g'=0.6$ that
allowes one to reproduce rather well the experimental energies of GTR in the nuclei in question.
The value $g'=1.0$ would correspond to the restoration of the SU(4) symmetry for p-h interaction
in our model.

The strengths of the monopole pairing particle-particle interaction $G_{0n},G_{0p}$ are chosen to
reproduce the experimental pairing energies in the considered nuclei.
The BCS basis has 9 levels (oscillator shells $N$=3,4) for $^{76}$Ge,
10 levels (oscillator shells $N$=3,4 plus the $1h_{11/2}$ orbit from $N=5$)
for $^{100}$Mo, and 12 levels (oscillator shells $N$=3,4 plus the $1h_{9/2,11/2}+2f_{7/2}$ orbits from $N=5$) for $^{130}$Te.
For such basis systems the realistic values are $G_{0n}\approx G_{0p}\approx 12$ MeV$/A$ with at most a variation of 20\%.
Hereafter, we will refer to this set of the pairing constants as set I.
We have checked that increasing the BCS basis by adding $N$=1,2 shells change the results only slightly.
Therefore, in addition, we choose artificially large $G_{0n}=G_{0p}=18$ MeV$/A$ (referred to as set II) as a test case in order
to emphasize some important points of our approach.

The strength of the triplet spin-spin particle-particle interaction $G_1$ is considered as a free parameter
and all dependencies are plotted as functions of the parameter $g'_{pp}=2G_1/(G_{0n}+G_{0p})$.
Note, that the basis for the QRPA diagonalization consists of all the bound s.p. states, thereby containing all
the main transitions among the spin-orbit doublets with the same radial quantum numbers.
The explicit representation for the set of the QRPA equations can be found elsewhere (see, e.g., \cite{Fae98}).
Note, that both $\beta^-$ and $\beta^+$ branches to construct the \bb-amplitude are calculated within the QRPA
in the same, initial for the \bb-decay, nucleus, so we adopt here the same approximation as in~\cite{Vog86}.

In Fig.~1 we show $S(g'_{pp})$ calculated according to the definition Eq.~(\ref{Mbb2}) (solid line) and its analytical
representation Eq.~(\ref{Sfin}) (dashed line). The horizontal dotted line corresponds to $S=0$.
Also the dependences corresponding to the pairing parameter sets I and II are plotted, where the latter is steeper due to the larger $\Delta_p,\Delta_n$.
In all cases the deviation of the numerical results from the analytical ones (with the largest deviation observed in $^{100}$Mo)
can barely be seen despite the fact that our QRPA bases do not contain the whole p-h spectrum as assumed in deriving Eq.~(\ref{Sfin}).
It gives one more confirmation that the QRPA is capable of reproducing well various sum rules.
Larger pairing correlations produce larger deviations due to a larger contribution of the p-h configurations discarded in the truncated basis.
It is noteworthy that the calculations of $S$ in $^{76}$Ge using the G-matrix generated residual interaction
have also revealed the linear dependence $S(g_{pp})$ with almost the same value $S(g_{pp}=0)$ (see Fig.~1 of~\cite{Civ00}).
It can also be seen from Fig.~1 that the realistic dependence $S(g'_{pp})$ is almost the same for different nuclei.
It can be understood from the phenomenological proportionalities $\Delta\propto A^{-1/2}$
and $G_0\propto A^{-1}$ put into $S\propto \Delta_n\Delta_p/G_0$.

\begin{figure}[thb]
 \begin{center}
\includegraphics[width=\textwidth,height=0.9\textheight]{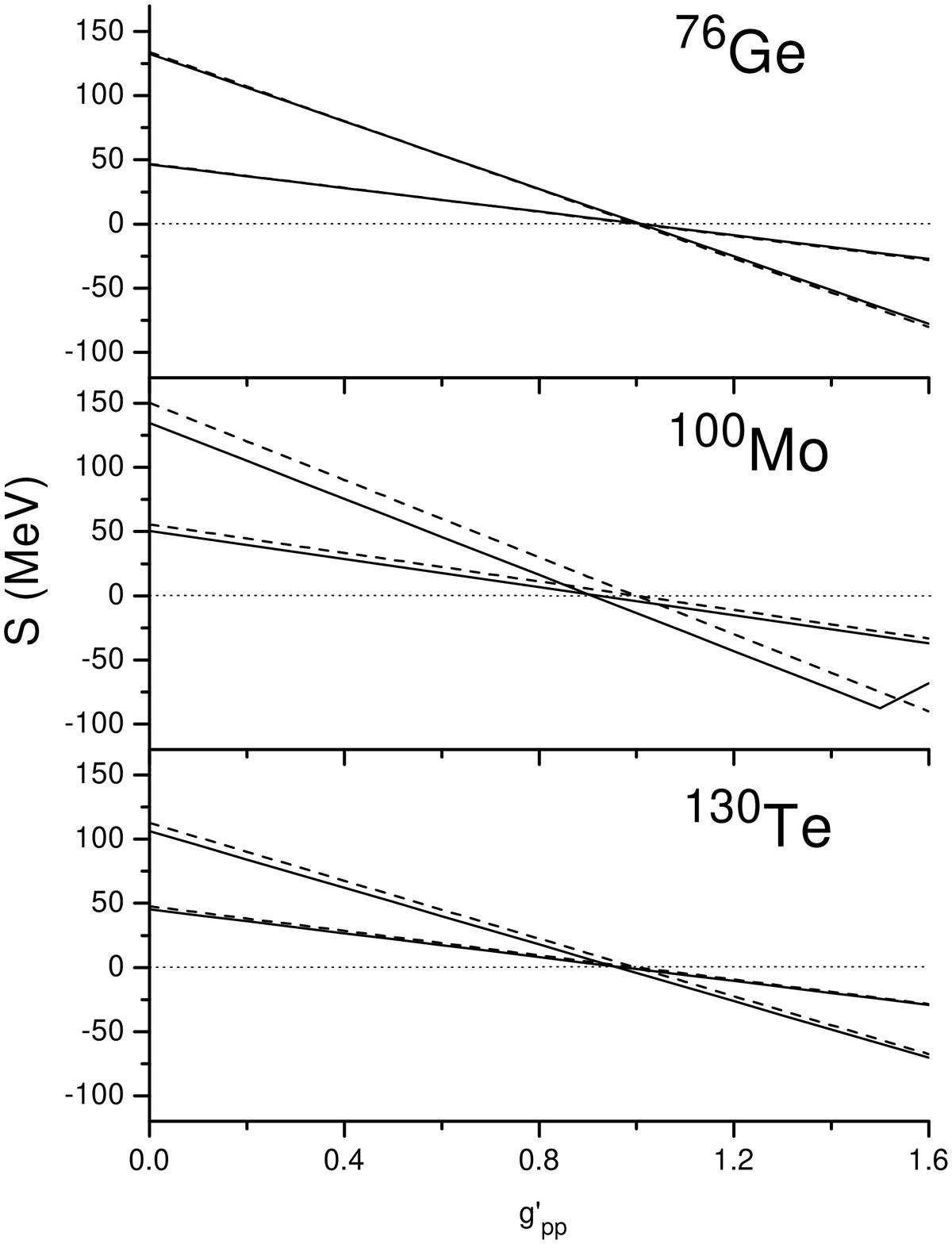}
\caption{The energy-weighted sum rule $S$ calculated according to the definition, Eq.~(\protect\ref{Mbb2}), (solid line) and its analytical
representation Eq.~(\protect\ref{Sfin}), (dashed line), for the pairing parameter sets I and II (the latter is steeper).}
 \end{center}
\end{figure}

In Fig.~2 we plot $M_{2\nu}(g'_{pp})$ (solid line) calculated according to Eq.~(\ref{Mbb}), $S/\omega_g^2$ (dashed line)
and their difference $M'_{2\nu}(g'_{pp})$ (dot-dashed line).
The value of $\omega_g$ is calculated as the mean energy of the Gamow-Teller strength distribution in the $\beta^-$-channel.
For each nucleus we have 
two sets of calculations - left and right panels contain the results corresponding to the pairing constant sets I and II, respectively.
One can see from the figure that the dependence $M'_{2\nu}(g'_{pp})$ is smoother than the original one, $M_{2\nu}(g'_{pp})$
(at least for not-too-large values of $g'_{pp}$ where the dependencies become essentially non-linear in the parameter).
Furthermore, by going to the unphysical set II of the pairing parameters, one can see that the main change in the amplitude of $M_{2\nu}(g'_{pp})$
can be attributed to the change of $S(g'_{pp})$ with a smaller change of $M'_{2\nu}(g'_{pp})$.
As for the function $S/\omega_g^2$, it shows basically a linear dependence on $g'_{pp}$ controlled by $S$.

\begin{figure}[h]
 \begin{center}
\includegraphics[width=0.5\textwidth,height=0.3\textheight]{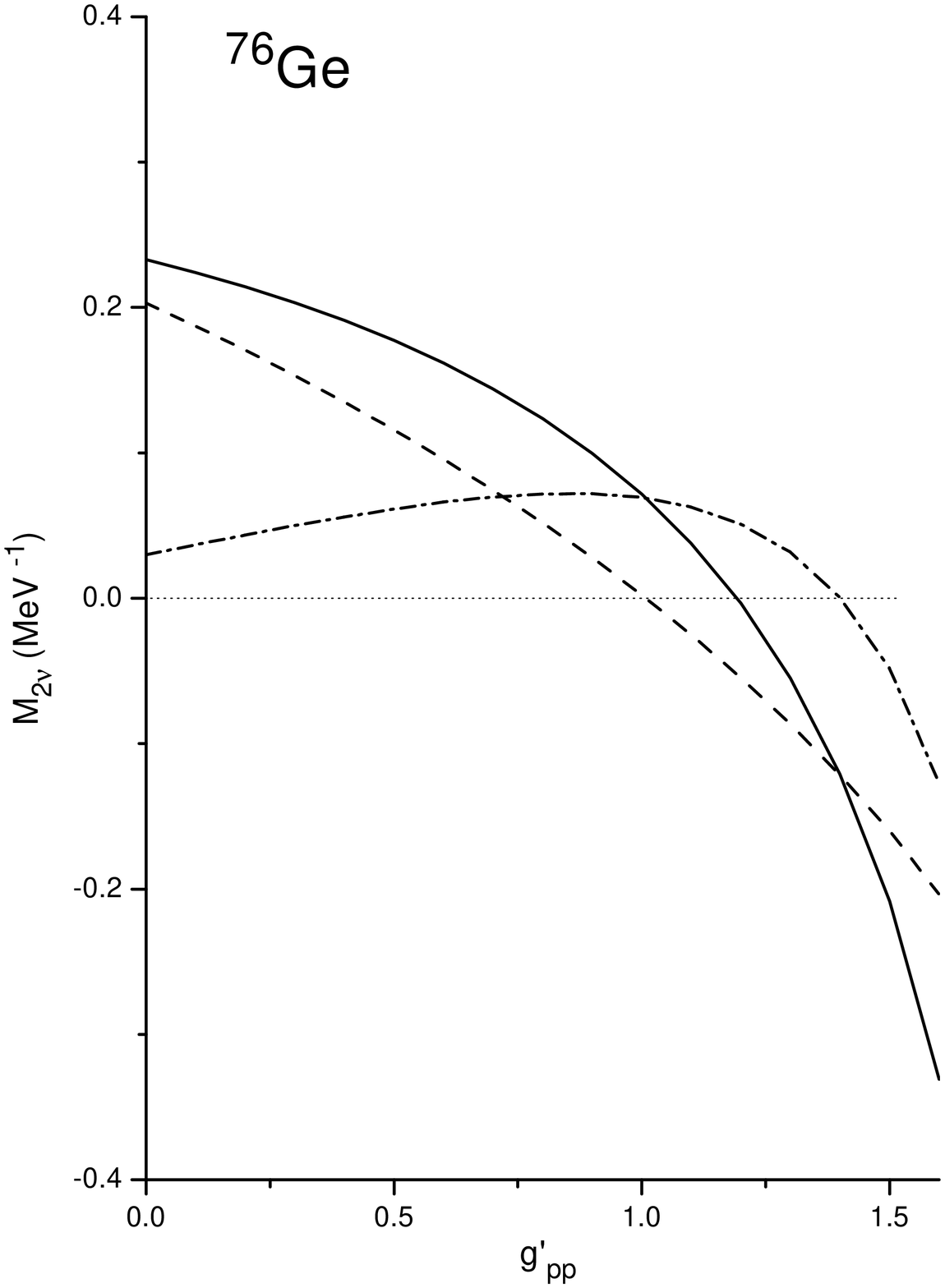}\includegraphics[width=0.5\textwidth,height=0.3\textheight]{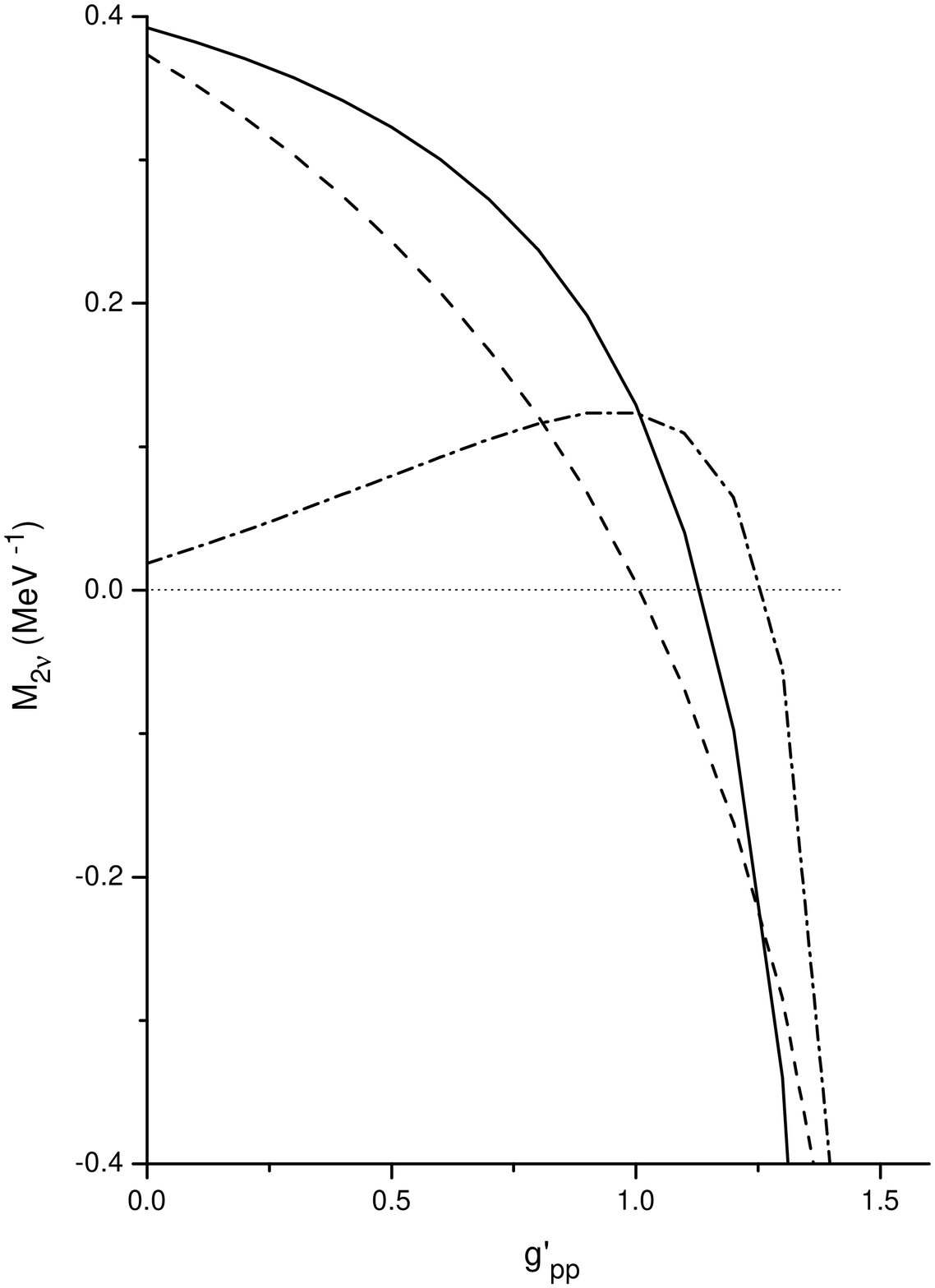}
\includegraphics[width=0.5\textwidth,height=0.3\textheight]{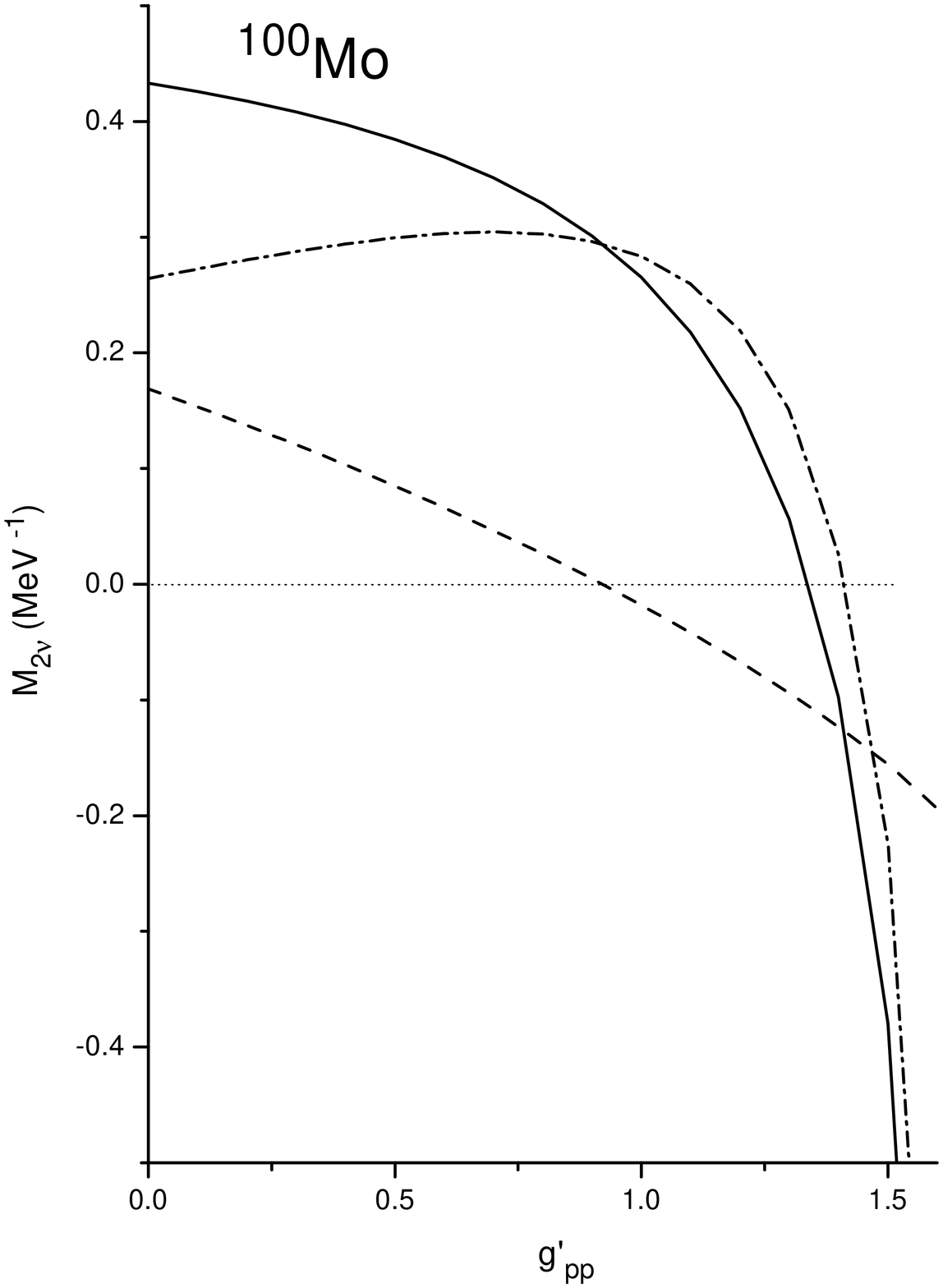}\includegraphics[width=0.5\textwidth,height=0.3\textheight]{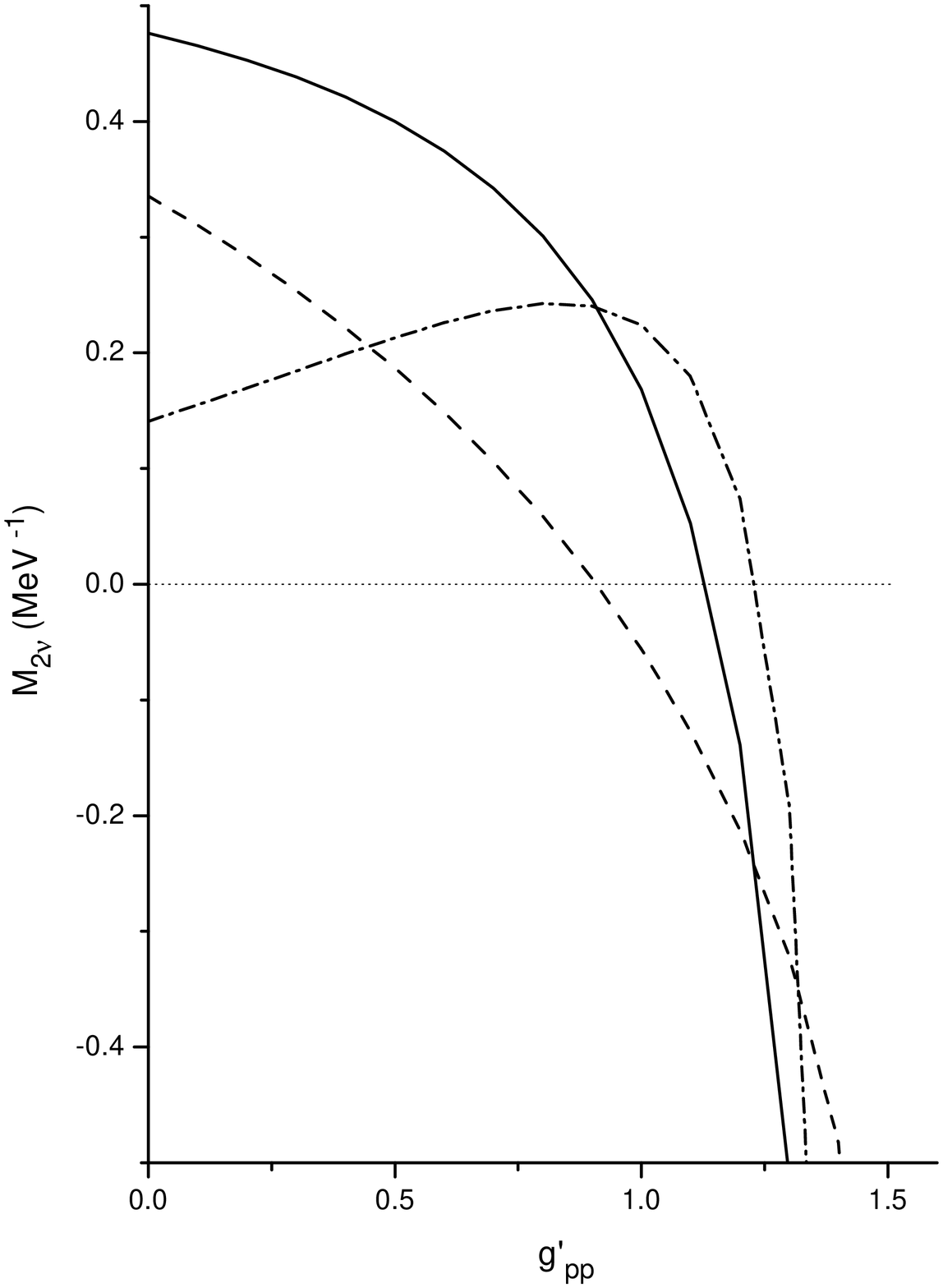}
\includegraphics[width=0.5\textwidth,height=0.3\textheight]{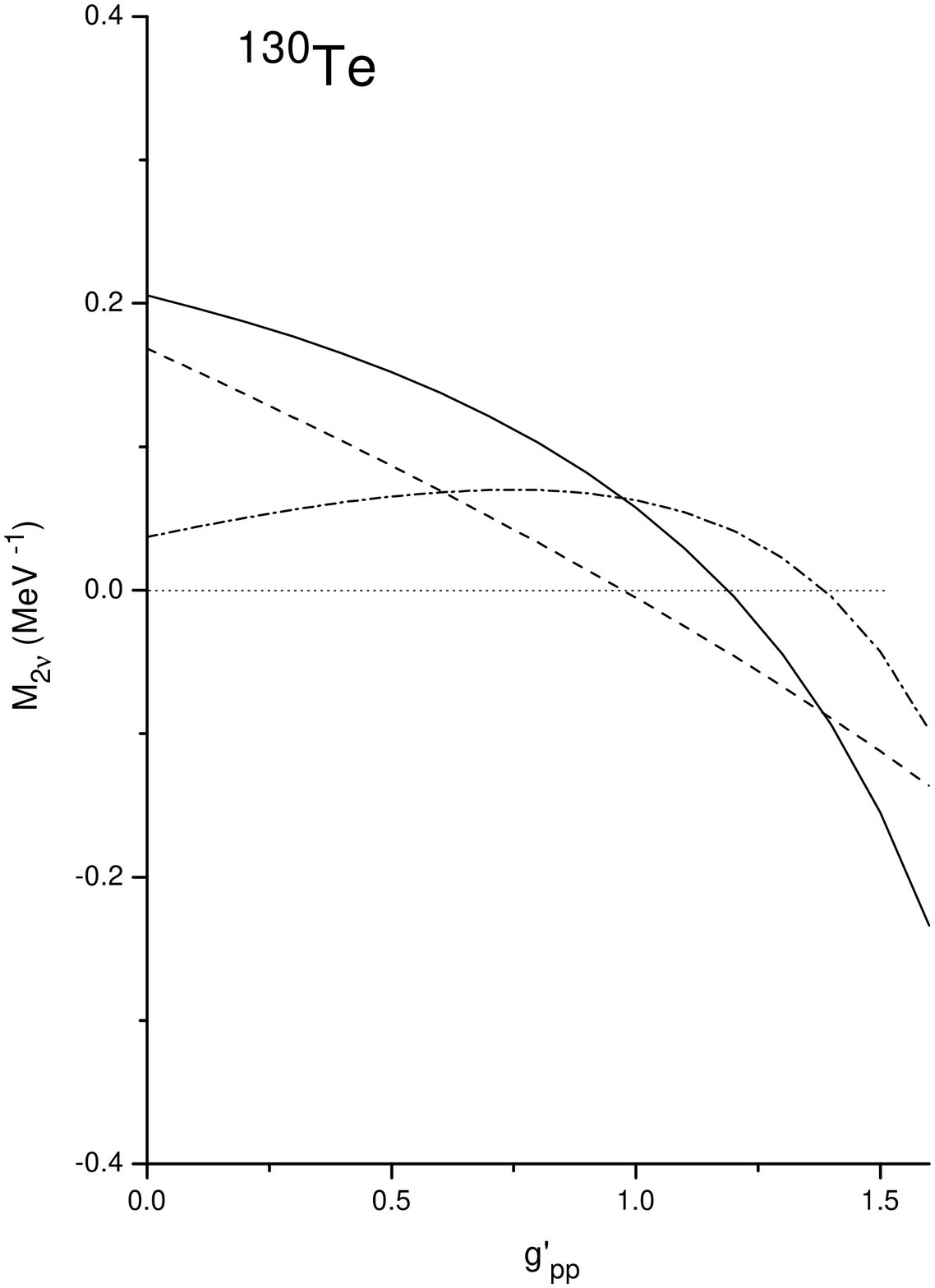}\includegraphics[width=0.5\textwidth,height=0.3\textheight]{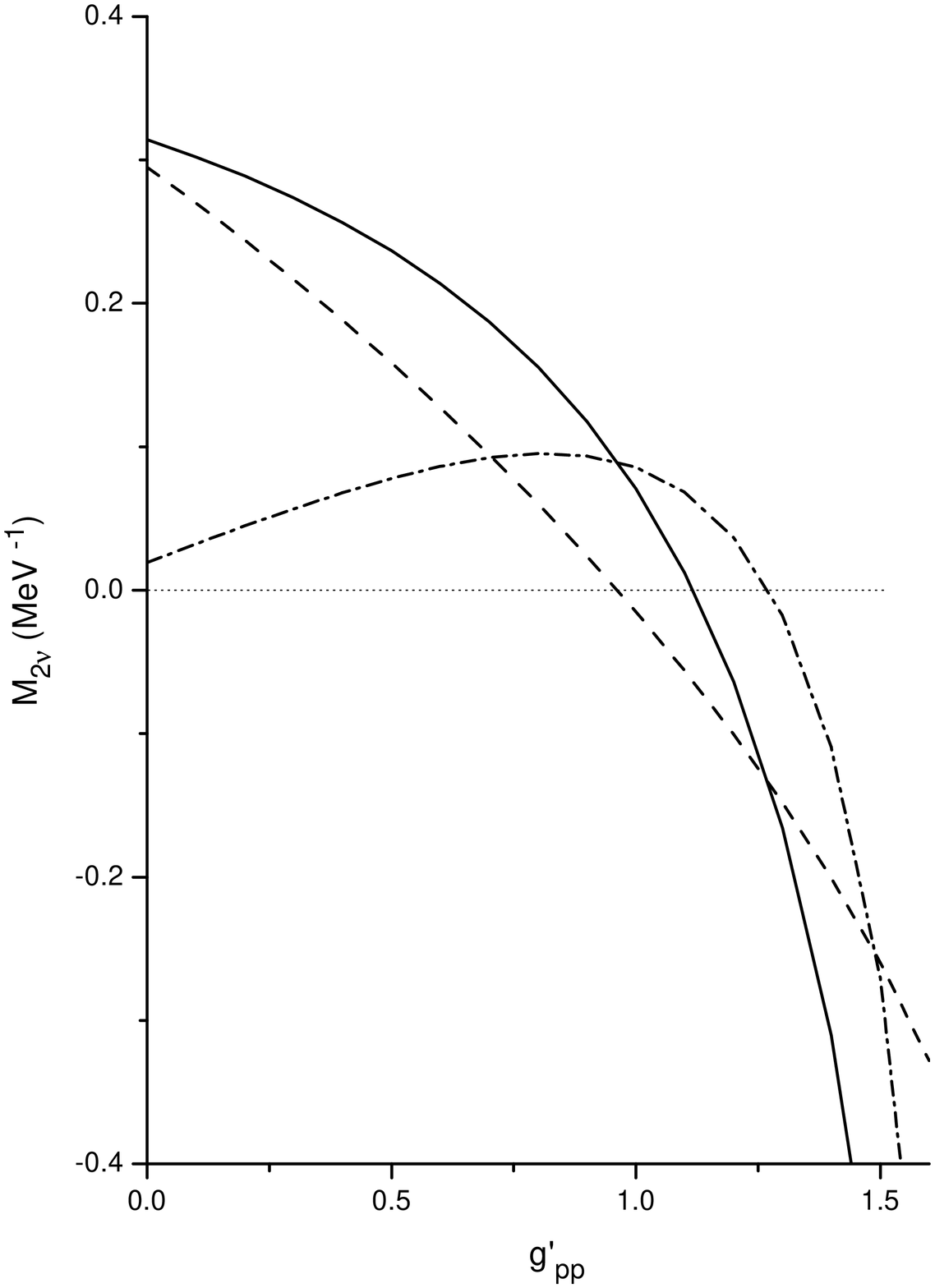}
\caption{The decomposition of the calculated $M_{2\nu}(g'_{pp})$ (solid line) as the sum of $S/\omega_g^2$ (dashed line) and
$M'_{2\nu}(g'_{pp})$ (dot-dashed line). The two sets of calculations for each nucleus are contained in the left and right panels
corresponding to the pairing constant sets I and II, respectively.
}
\end{center}
\end{figure}

Finally, we have tested the main conjecture of~\cite{Rum98} that  $M'_{2\nu}$ (\ref{M_G'}) is mainly determined by the spin-orbit mean field contribution
to the commutator (\ref{VG}). We plot the corresponding results
in Fig.~3.
The solid and dash-dotted lines represent $M'_{2\nu}(g'_{pp})$ calculated according to the
definition (\ref{Mbb'}) for 100\% and 50\% of the realistic strength of the spin-orbit interaction, respectively.
The dashed and dotted lines represent $M'_{2\nu}(g'_{pp})$ calculated according to the (\ref{M_G'}),(\ref{VG}), taking only spin-orbit
contribution to the latter equation (see~\cite{Rum98} for the details), also for 100\% and 50\% of the realistic strength of the spin-orbit interaction,
respectively.
One can see from Fig.~3 that $M'_{2\nu}(g'_{pp})$ is indeed very sensitive to the change of $U_{so}$.
The approximation of $M'_{2\nu}(g'_{pp})$
by (\ref{M_G'}),(\ref{VG}) is fairly good for the realistic $g'_{pp}$ in the vicinity of unity
(a recent discussion of the realistic $g_{pp}$ can be found in~\cite{Rod03a}). Thus, in that region $M'_{2\nu}(g'_{pp})$ should scale
almost quadratically with the magnitude of $U_{so}$ and the realistic choice of the spin-orbit interaction intensity is very important.
It is also noteworthy that the schematic model QRPA calculations~\cite{Vog86} revealed almost linear dependence of the \bb-amplitude on $g'_{pp}$ that
is quite understandable because the authors neglected the spin-orbit splitting in the
model\footnote{unfortunately, it is not possible to compare our results with the values of the \bb-amplitude from Fig.~2 of~\protect\cite{Vog86}
because the ordinate was given in arbitrary units.}.
\begin{figure}[thb]
 \begin{center}
\includegraphics[width=\textwidth,height=0.9\textheight]{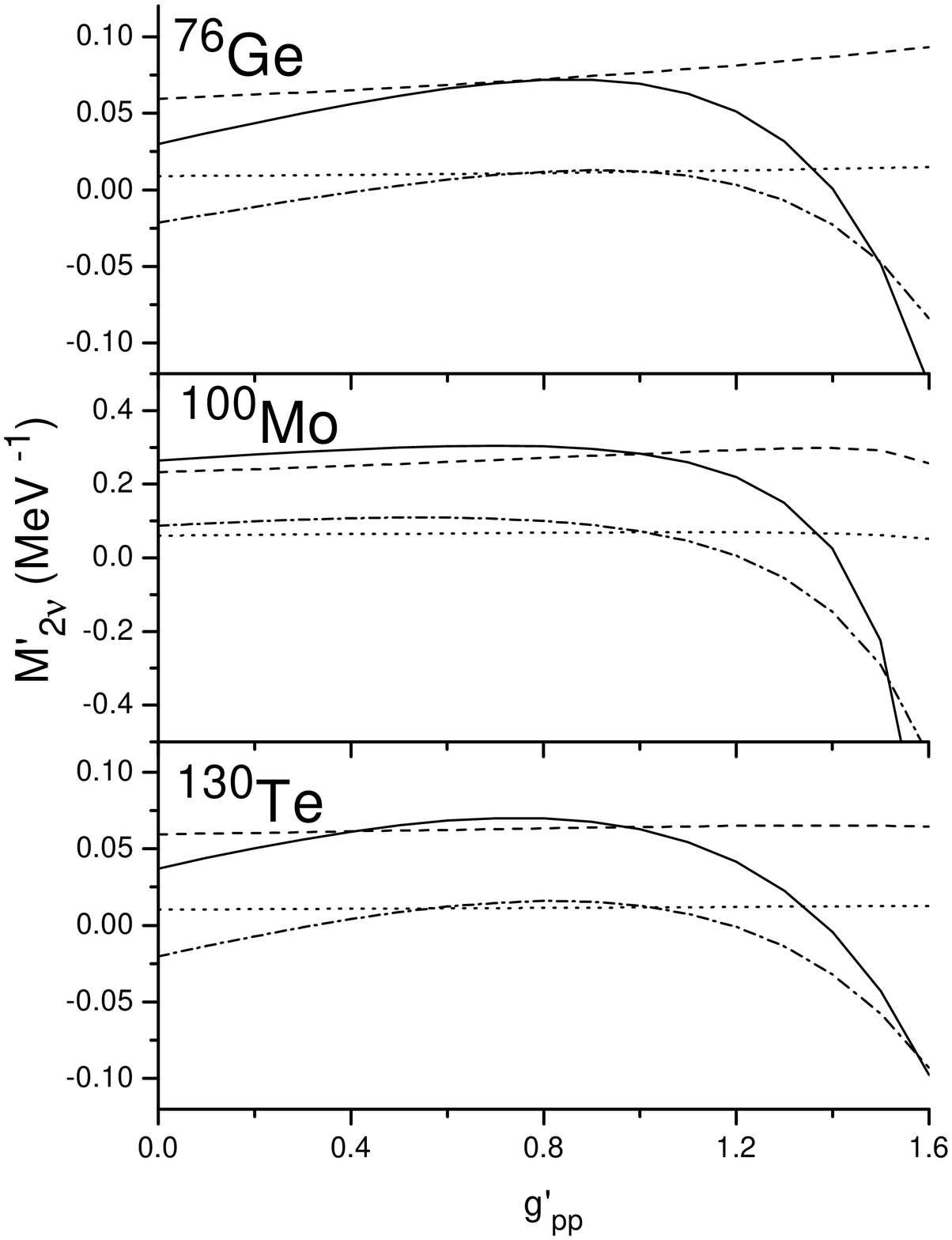}
\caption{$M'_{2\nu}$ calculated according to the
definition (\protect\ref{Mbb'}) for 100\% (solid line) and 50\% (dash-dotted line) of the realistic strength of the spin-orbit interaction.
The dashed and dotted lines represent $M'_{2\nu}$ calculated according to the (\protect\ref{M_G'}),(\protect\ref{VG}), taking only the spin-orbit
contribution to the latter equation for 100\% and 50\% of the strength of the spin-orbit interaction,
respectively.}
 \end{center}
\end{figure}

\section{Conclusions}

In this paper, using an identity transformation, we have represented the \bb-amplitude as a sum of two terms.
One term has been shown to be proportional to an specific energy-weighted sum rule depending only on the
particle-particle part of the Hamiltonian. This part has been shown within the QRPA to account for most of the sensitivity of the original {\bb}-amplitude to $g'_{pp}$ for realistic $g'_{pp}\simeq 1$. 
The sum rule has been found within the QRPA to depend linearly on the ratio $g'_{pp}$
and to vanish at the point $g'_{pp}=1$ where the Wigner SU(4) symmetry is restored in the p-p sector of the Hamiltonian.
The analytical results for the sum rule have been reproduced very well by the QRPA calculations using truncated p-h bases.
Thus, we have shown that the sensitivity of the amplitude to $g'_{pp}$ is due to the pronounced effect of
violation of the SU(4) symmetry in the p-p sector of a realistic nuclear Hamiltonian.
The second term, $M'_{2\nu}$, in the representation of
the {\bb}-amplitude has been shown within the QRPA to be a much smoother function for realistic values of $g'_{pp}$ 
than the original \bb-amplitude.
We have confirmed the conjecture of Rumyantsev and Urin~\cite{Rum98} that $M'_{2\nu}$ is mainly determined by the intensity of the spin-orbit interaction of the nuclear mean field
and scales approximately as square of the intensity for the realistic values of $g'_{pp}$.
Therefore, the realistic choice of the spin-orbit mean field is very important for describing the {\bb}-amplitude.
Finally, the analysis of the present work has revealed the reasons for 
the sensitivity of the {\bb}-amplitude to different components of the nuclear Hamiltonian and thereby can help in constraining nuclear model uncertainties in calculations of the amplitude.

\section*{Acknowledgments}
This work was supported in part by Landes\-forschungs\-schwerpunkts\-programm
Baden-W\"urttemberg ``Low Energy Neutrino Physics".
V.A.R. would like to thank the Graduiertenkolleg ``Hadronen im Vakuum,
in Kernen und Sternen" (GRK683 of DFG) for support.
M.H.U. acknowledges financial support from the ``Nederlandse organisatie voor wetenschappelijk
onderzoek" (NWO) during his stay at the KVI.

\end{document}